\documentclass[5p]{elsarticle}

\usepackage{hyperref}
\usepackage{epsfig, bm, amsfonts, amssymb}
\usepackage[normalem]{ulem}
\usepackage{amsmath}
\usepackage{flushend}
\usepackage{cuted}
\usepackage{breqn}
\RequirePackage{color}

\definecolor{MyDarkGreen}{rgb}{0.02,0.60,0.06}

\usepackage{epstopdf}

\journal{Physics Letters A}

\begin{document}
\begin{frontmatter}

    \title{Ising model with invisible states on scale-free networks}

    \author[A,B,C]{Petro Sarkanych\fnref{myfootnote}}
    \author[A,C]{Mariana Krasnytska}
    \address[A]{~Institute for Condensed Matter Physics, National Academy of Sciences of Ukraine, Lviv, Ukraine}
    \address[B]{~Centre for Fluid and Complex Systems, Coventry University, Coventry, UK}
    \address[C]{~${\mathbb L}^4$ Collaboration \& Doctoral College for the Statistical Physics of Complex Systems, Leipzig-Lorraine-Lviv-Coventry, Europe}

    \fntext[myfootnote]{sarkanyp@uni.coventry.ac.uk}

    \begin{abstract}
        We consider the Ising model with invisible states on scale-free networks. Our goal is to investigate the interplay between the entropic and topological influence on a phase transition. The former is manifest through the number of invisible states $r$, while the latter is controlled by the network node-degree distribution decay exponent $\lambda$. We show that the phase diagram in this case is characterised by two marginal values
        $r_{c1}(\lambda)$ and $r_{c2}(\lambda)$, which separate regions with different critical behaviours. Below the $r_{c1}(\lambda)$
        line the system undergoes only second order phase transition; above the $r_{c2}(\lambda)$ - only a first order phase transition occurs;
        and in-between the lines both of these phase transitions occur at different temperatures. This behaviour differs from the one,
        observed on the lattice, where the Ising model with invisible states is only characterised with one marginal value $r_{c}\simeq 3.62$
        separating the first and second order regimes.
    \end{abstract}


\end{frontmatter}

\section{Introduction}
\label{I}

Many inherent features of phase transitions in many particle systems can
be understood and quantitatively described by the analysis of an
interplay between entropy and energy. Taking magnetic ordering as an
example, energy-entropy interplay allows one to explain the absence of spontaneous
magnetisation at low dimensions \cite{Ruelle1968,Landau} or the influence of
structural (topological) disorder on magnetic  ordering\cite{zittartz_1,zittartz_2,krasnytska1,Krasnytska13}.
Therefore, much attention has been paid to the analysis of ordering phenomena in
many-particle systems by introducing models that give direct access to
trigger the system's entropy in a countable way. One of them is the recently introduced
Potts model with invisible states \cite{Tamura10,Tamura11,Tanaka11a}.
Unlike standard $q-$state Potts model, this modification possesses additional
$r$ invisible states. If a spin lies in one of these invisible states, it
does not interact with the rest of the system. Thus introducing invisible
states does not change the interaction energy, but rather the number of
configurations, or equivalently - the entropy. This model was originally suggested
to explain why the phase transition with the
$q-$fold symmetry breaking undergoes a different order than
predicted theoretically \cite{Tamura10,Tamura11,Tanaka11a}.

Analysis of this model on different lattices has been a subject of intensive
analytic \cite{Johnston13,Mori12,Enter11a,Enter11b,Ananikian13,Sarkanych17,Sarkanych18} and numerical \cite{Tamura10,Tamura11,Tanaka11a}
studies. It has been shown that the
number of invisible states ($r$)  plays the role of a parameter, whose
increase makes the phase transition sharper. For example the $q=2$ model
with $r=30$ invisible states on a square lattice undergoes a first order
phase transition, while $q=2$ and $r=0$ correspond to the ordinary Ising
model, which is a textbook example of a second order phase transition \cite{Tamura10}.

Interesting phenomena were observed for the Potts model with invisible
states when it is considered on a complete graph \cite{Krasnytska16}.
In the region $1\leq q< 2$ it
possesses untrivial critical behaviour \cite{Krasnytska16}: for
small values of $r$ the system undergoes only second order phase
transition. For large $r$ there is only a first order
phase transition. In-between there is a region where both phase
transitions occur at different temperatures. Thus, the phase diagram is
characterised by two marginal values: $r_{c1}$ - where the first order phase transition appears, 
and $r_{c2}$ - where the second order phase transition disappears. In the Ising case $q=2$
the phase diagram is characterised by one critical value $r_c\simeq3.62$, which separates
regions with first and second order phase transitions.

In this paper we consider the above described model on a complex network \cite{networks_1}-\cite{networks_5},
being primarily interested in the Ising case $q=2$. Much attention has been paid
to the study of phase transitions on complex networks \cite{Dorogovtsev08}. Besides pure academic interest,
such problems have a number of practical motivations, ranging from sociophysics,
where the structure of social interations is properly described by a network topology
to nanophysics, where network reflects structure of particle agregates.
Of particular interest are the scale-free networks, where the node-degree distribution
(the probability of a randomly chosen vertex to have degree $k$)
is governed by a power-law decay:
\begin{equation}
\label{pofk}
P(k)\sim 1 /{k^\lambda}, \, k \to \infty \, .
\end{equation}
It has been shown that many standard models of statistical physics
manifest unusual features when considered on scale-free
networks
\cite{Krasnytska13,Leone02,Dorogovtsev02,Palchykov10,Igloi02}. In
particular, it has been found that the decay exponent $\lambda$
determines the collective behaviour \cite{Leone02,Dorogovtsev02} and
its continuous change plays a similar role as space
dimensionality for lattice systems
\cite{Holovatch92,Holovatch98}. In particular, the Ising model on a
scale-free network is characterized by the lower and upper critical
values of $\lambda$: below  $\lambda = 3$ the system is ordered at
any finite temperature, above $\lambda = 5$ the system is governed
by the usual mean field critical exponents, whereas in the intermediate
region the critical exponents become $\lambda$-dependent
\cite{Dorogovtsev08,Leone02,Dorogovtsev02}. Moreover, logarithmic
corrections to scaling appear at $\lambda=5$ \cite{Palchykov10}. In
turn, for the standard $q$-state Potts model on a scale free network the
values of $\lambda$ and $q$ determine the order of phase transition
\cite{Krasnytska13,Igloi02,Krasnytska14}.

Being introduced rather recently, the Potts model with
invisible states has not yet been a subject of analysis on a
scale-free network. Although its analysis on a complete graph brings
about rather unexpected critical behaviour \cite{Krasnytska16}.
Therefore, it is tempting to perform such a study to analyse the
combined impact of different factors that allow to trigger an amount
of disorder in many-particle system. Moreover, considering the model
with invisible states on a scale free network allows one to study within
the unique approach an interplay of different forms of
disorder: one arising from the number of configurations of
the internal degrees of freedom (number of invisible states $r$) and
another one arising from structural inhomogeneities `hubs' (node
degree distribution exponent $\lambda$).

The rest of the paper is organised as follows: in Section \ref{II} we apply mean-field approach
to find the free energy; having this result to hand we proceed with numerical analysis in Section \ref{III};
we draw conclusions in Section \ref{IV}.

\section{Model and mean-field approximation}
\label{II} The Hamiltonian for the Potts model with invisible states
reads, see \cite{Krasnytska16}:
\begin{equation}\label{1a}
- H(q,r)=\sum_{<i,j>}J _{ij}\sum_{\alpha=1}^q \delta
_{S_i,\alpha}\delta _{\alpha,S_j}+ h\, \sum _{i=1}^N  \delta_{S_i,1},
\end{equation}
where $s_i=(1,...,q,q+ 1,...,q+r)$ is  the  Potts  variable, $q$ and $r$ are  the  numbers  of  visible  and  invisible  states
respectively,  $\delta_{\alpha,S_j}$ is  Kronecker  delta  symbol and an external magnetic field $h$ is introduced
to favour the first visible state. The first summation in (\ref{1a}) is performed over all pairs of
spins in the network of $N$ nodes, the second sum requires both of the interacting spins to be in the same visible state.
Considering this model on the network, one assumes the couplings $J_{ij}$ to be in the form of network
adjacency matrix: $J_{ij}=1$ if node $i$ and $j$ are connected and $J_{ij}=0$ otherwise.

All further analytical results will be performed for general $q$ and numerically evaluated for
the Ising case $q=2$. For the purpose of our analysis we adopt a variant of the mean-field approach presented in
Refs. \cite{Krasnytska13,Igloi02}.
Let us introduce the local thermodynamic averages:
\begin{eqnarray}\label{2'}
\langle \delta
_{S_i,\alpha} \rangle=\left\{
\begin{array}{ccc}
& \mu_i \, , & \alpha=1, \\
& \nu_{1i} \, , & \alpha=2,..q, \\
& \nu_{2i}\, , & \alpha=q+1,...r\, ,
\end{array}
\right.
\end{eqnarray}
where the thermodynamic averaging is performed with the Hamiltonian (\ref{1a}). The normalization condition
\begin{equation}\label{3'}
\mu_i+(q-1)\nu_{1i}+r\nu_{2i}=1
\end{equation}
enables one to construct two independent local order parameters. Taking into account low- and high-temperature asymptotics of
the averages (\ref{2'}) and (desired) asymptotics for the order parameters, see Table \ref{tab1}, one can define two local
order parameters by
\begin{equation}\label{4'}
m_{1i}=\mu_i-\nu_{1i}, \quad
m_{2i}=\mu_i-\nu_{2i}.
\end{equation}

\begin{table}[b]
    \caption{Low and high temperature asymptotics of the thermodynamic
        averages, Eq.~(\ref{2'}), and order parameters, Eq. (6).
        \label{tab1}}
    \begin{center}
        \begin{tabular}{|c|c|c|c|c|c|}
            \hline
            $\beta \rightarrow \infty$ & $\mu=1$ & $\nu_{1}=0$ & $\nu_{1}=0$ & $m_1=1$ & $m_2=1$ \\  \hline
            \hline
            $\beta\rightarrow 0$ & $\mu=\frac{1}{q+r}$ & $\nu_{1}=\frac{1}{q+r}$ & $\nu_{1}=\frac{1}{q+r}$ &
            $m_1=0$ & $m_2=0$ \\
            \hline
        \end{tabular}
    \end{center}
\end{table}


Using definitions for the local averages (\ref{2'})
and neglecting second order contributions from the
fluctuations $\delta
_{S_i,\alpha} - \langle \delta
_{S_i,\alpha} \rangle$ we arrive at the mean-field Hamiltonian:
\begin{eqnarray}\label{9'}
- H(q,r)=\sum_{<i,j>}J_{ij}[\mu_i(2\delta
_{1,S_j}-\mu_j)+\\
\nonumber \sum_{\alpha=2}^q (2\delta
_{\alpha,S_i}-\nu_{1i})\nu_{1j}]+h\sum_i \delta_{S_i,1}.
\end{eqnarray}
Taking trace of (\ref{9'}) over all possible
configurations of spins we obtain free energy per site:
\begin{eqnarray}
\label{f'} f(\mu,\nu_1)=\sum_j
J_{ij}(\mu_i\mu_j+(q-1)\nu_{1i}\nu_{1j})-\\
\nonumber \frac{1}{\beta}\sum_i
\ln\Big( e^{\beta(h+2\sum_j J_{ij}\mu_j)}+(q-1)e^{2\beta\sum_j
    J_{ij}\nu_{1j}}+r\Big).
\end{eqnarray}

Within the mean-field approach we also consider a coupling constant proportional to the probability of two nodes being connected
\begin{equation}\label{11a}
J_{ij}=Jp_{ij}=\frac{J k_i k_j}{N\bar{k}},
\end{equation}
where $k_i$ stands for the degree of node $i$ and $\bar{k}$ is average node degree in the network. We also introduce a global weighted order parameters according to the rules
\begin{equation}\label{11b}
m_1=\frac{\sum_i k_i m_{1i}}{\sum_i k_i}, \hspace{2cm}
m_2=\frac{\sum_i k_i m_{2i}}{\sum_i k_i}.
\end{equation}

With all these substitutions in the thermodynamic the limit free energy per site as a function of global order parameters reads
\begin{dmath}
\label{ff0}
f(m_1,m_2)=\frac{J\langle k\rangle}{(q+r)^2}\Big((rm_2+1+(q-1)m_1)^2+
(q-1)(rm_2+1-(r+1)m_1)^2\Big)-\frac{1}{\beta}\int_2^\infty dk P(k) \ln\Big(e^{\beta(h+\frac{kJ}{q+r}(m_1(q-1)+1+rm_2))}+(q-1)e^{\frac{\beta Jk}{q+r}(m_2r+1-(r+1)m_1)}+r\Big),
\end{dmath}
where $P(k)$ is a node degree distribution. Henceforth we will only consider a scale-free network that is governed
by a power law decay (\ref{pofk}).

Eq. (\ref{ff0}) gives the free energy as a function of two order parameters $m_1$ and $m_2$ with a set of
parameters $q,r,\beta,\lambda$. We will also set $J=1$ so the temperature is measured in these energy units.
Usually in such a case, the next step is to present the free energy as a power series over order parameters
(Landau free energy). In our case two order parameters make this expansion too cumbersome for a direct analytic treatment 
and we switch to numerical analysis of the free energy. For this purpose we adopt simplex method \cite{NelderMead}. Its advantage
is that it does not require to know the derivatives of the function and only needs a way to evaluate it. With this
numerical technique to hand in the following Section we proceed according to the 
following scheme. For fixed values of $q, r$
and $\lambda$ we sweep through a certain region of temperatures and calculate the values of $m_1$ and $m_2$ which minimise
the free energy. Based on the temperature behaviour of the order parameters we can 
then make conclusions about
the order of the phase transition, critical temperature and critical exponents.

\section{Results}
\label{III} In this section we investigate the Ising model
$(q=2)$ with an arbitrary number of invisible states $r$ near the
spontaneous phase transition point ($h=0$) on a scale-free network.
All our results will be compared with the analytic results
known in the limit $r=0$ \cite{Leone02,Dorogovtsev02,Palchykov10}.
This particular case of the Ising model on a scale-free
network hereafter is called the "genuine" Ising model. We will mostly
be interested in the region $3\leq \lambda\leq5$, where $\lambda$
dependent critical exponents were observed.

For the invisible states Ising model on a scale-free
network, one would expect that the node degree distribution
exponent has a similar effect as for the genuine Ising model. Indeed,
our analysis supports this conjecture. In particular, for low values
of $\lambda\leq 3$ the system remains ordered at any finite
temperature. However, the region $\lambda\geq 3$ appears to exhibit some
not-trivial features we will discuss in more details below.

Let us start with an analysis of the critical temperature $T_c$. At critical temperature any ordering disappears. This is found when $m_1=0$. Earlier it was shown, that on a complete graph $m_2$ vanishes only at infinite temperature,
thus only the first order parameter can be used to determine the order of the phase transitions and the critical temperature.

\begin{figure}
    \includegraphics[width=\columnwidth]{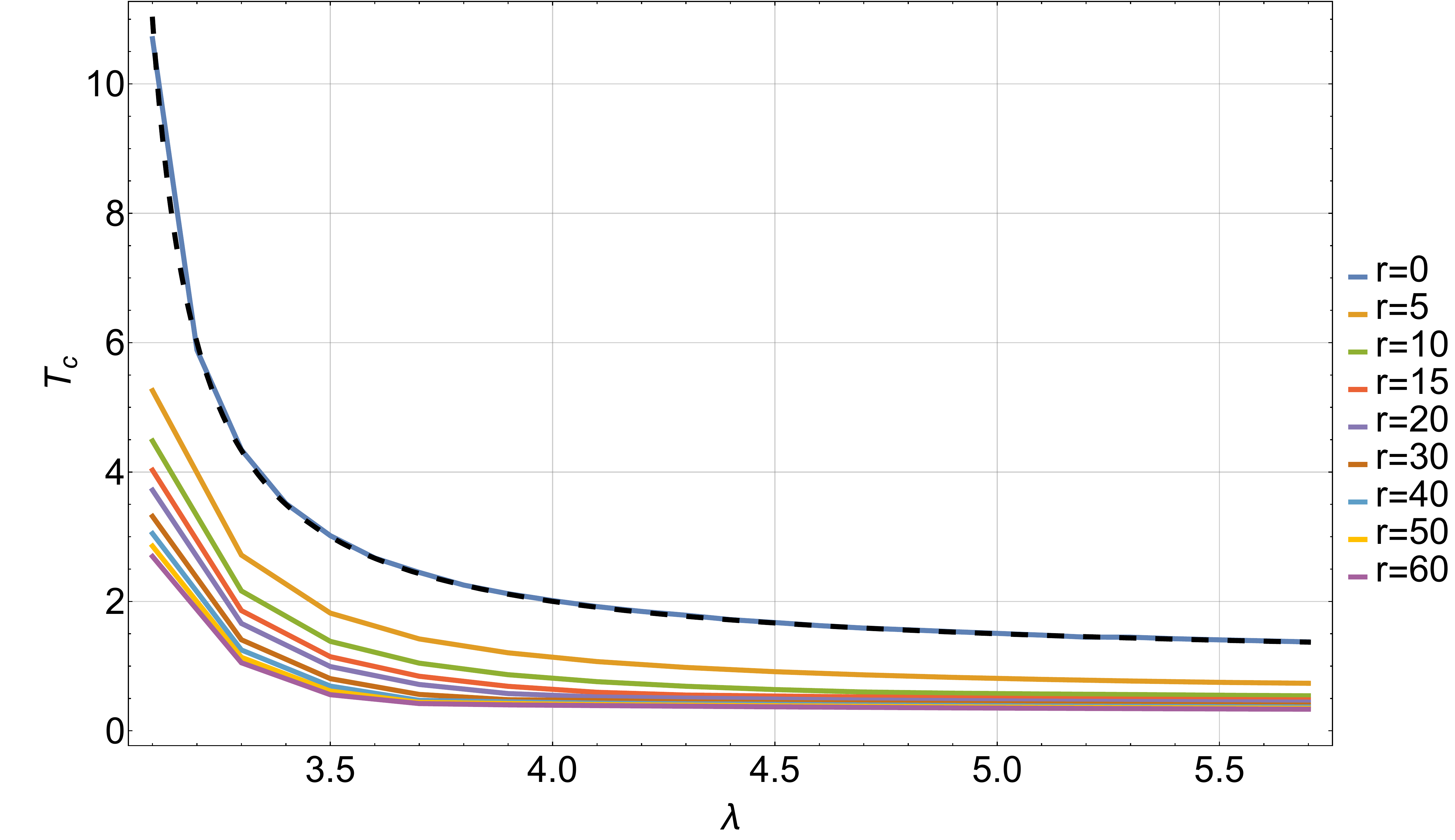}
    \caption{\label{fig1} Critical temperature of the Ising model with invisible states on a scale free network as a function of
    the degree distribution exponent $\lambda$ for different values of $r$: $r=0,5,10,15,20,30,40,50,60$ going down the plot.
    Dashed line represents analytical results for the genuine Ising case \cite{Leone02,Dorogovtsev02,Palchykov10}.}
\end{figure}

In Fig. \ref{fig1} the critical temperature of the Ising model with invisible states on a scale free network is given as a function of
$\lambda$ for various numbers of invisible states $r$ ranging from 0 to 60. Critical temperatures obtained with our numerical
technique are in a good agreement with analytical results for the genuine Ising model (see the upper solid and dashed lines in Fig. \ref{fig1}) \cite{Dorogovtsev02}.

From the plot, it is clear
that the critical temperature decreases with an increase of $\lambda$.
When $\lambda$ decreases below the marginal value
$\lambda=3$, no finite temperature can break spontaneous ordering:
the system remains ordered at any $T$. This reflects the fact, that
for small $\lambda$ there are many nodes with high degree (hubs),
making the network strongly connected. Taking this into account, in
the limit $\lambda\to 3+0$, critical temperature rises to
$T_c\to\infty$.

On the other hand, from Fig. \ref{fig1} one can also say, that the
critical temperature decreases with an increase in the number of
invisible states. This is because $r$ regulates the entropy
of the system, meaning that the more entropy there is, the easier it is to 
break the ordering. Limit $r\to\infty$ will reproduce
results for non-interacting system, i.e. $T_c=0$
\cite{Krasnytska16}.

\begin{figure}
    \includegraphics[width=\columnwidth]{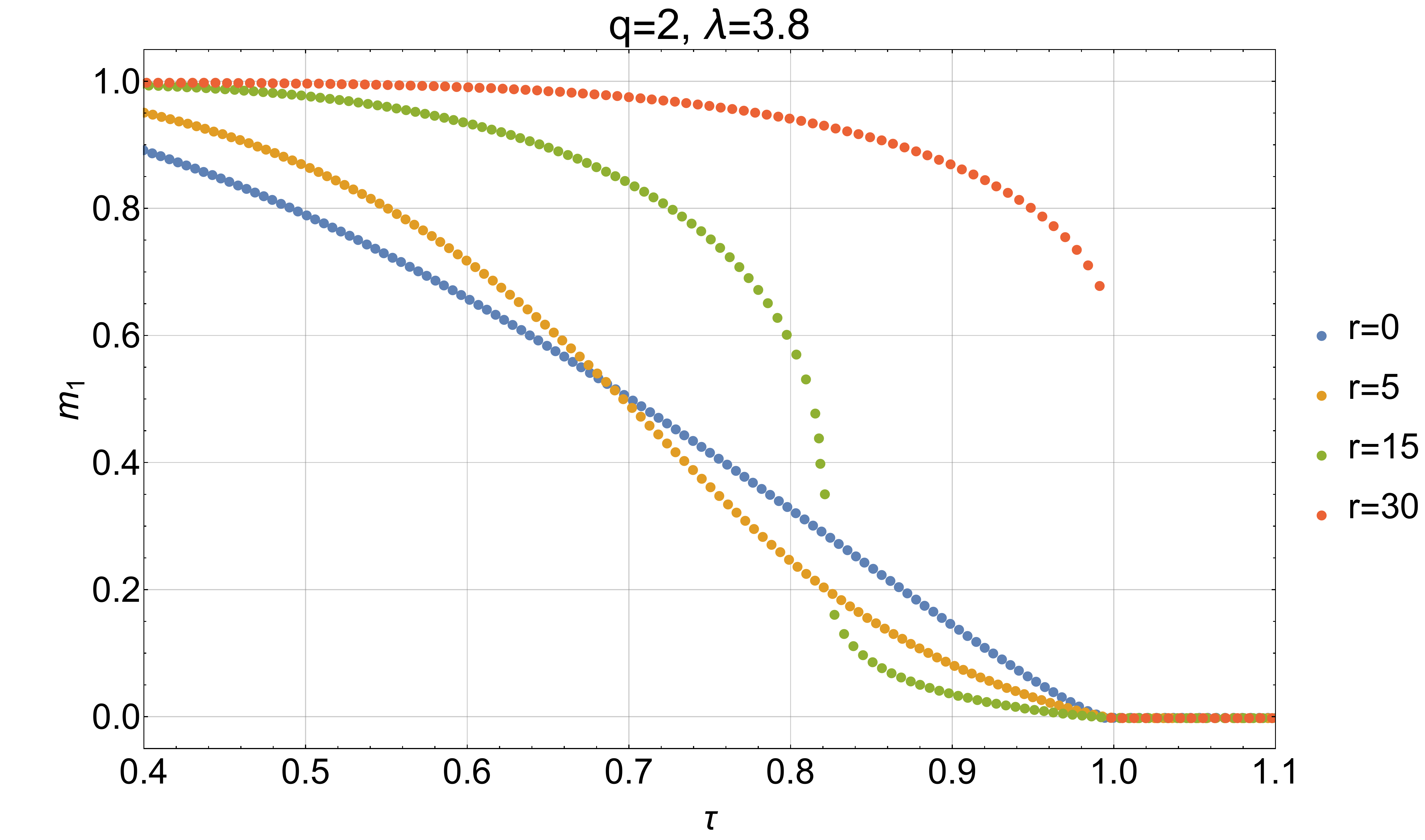}
    \includegraphics[width=\columnwidth]{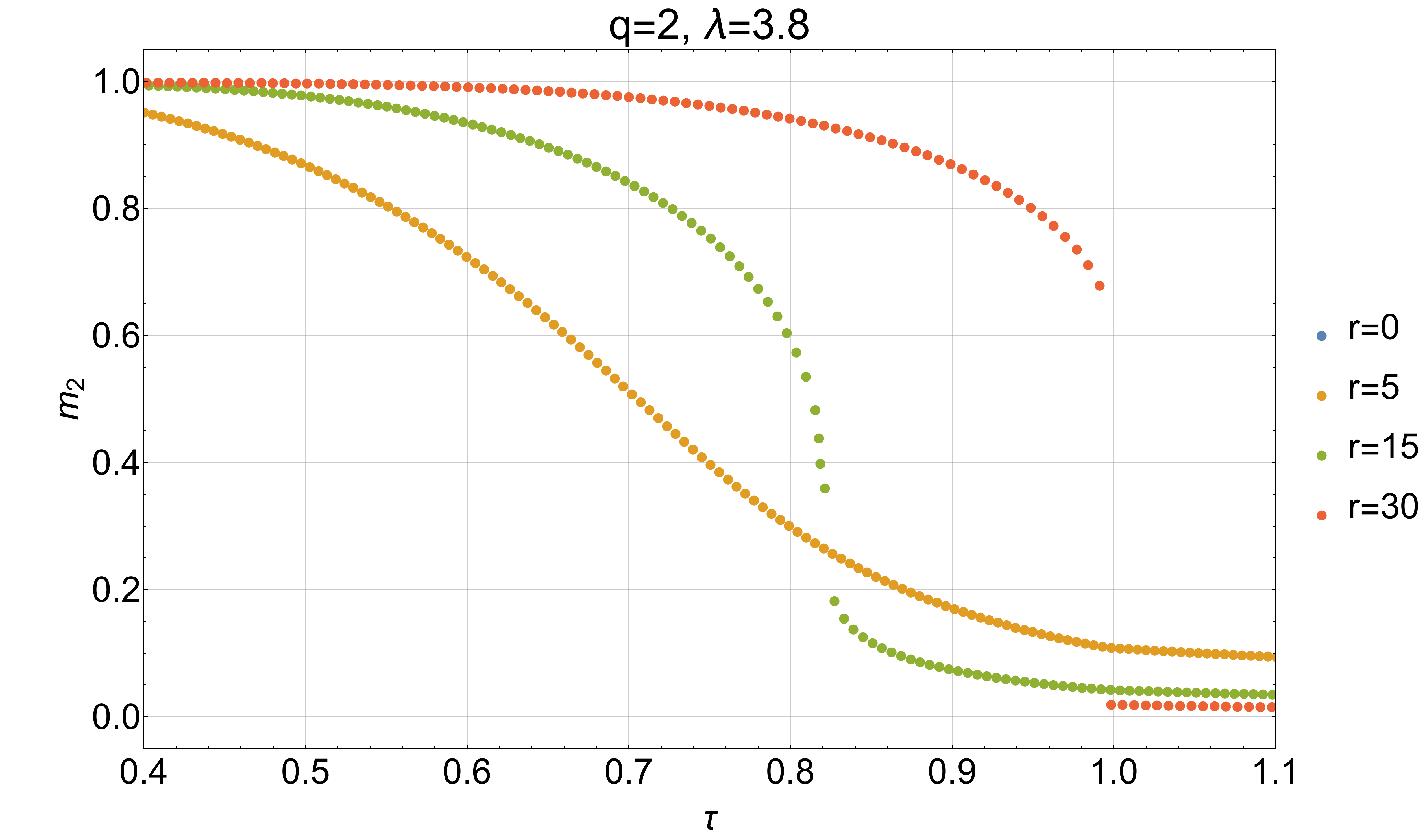}
    \caption{\label{fig3} Order parameters as functions of reduced temperature $\tau=T/T_c$ for various values of $r$ and fixed $\lambda=3.8$.
    Different values of $r$ lead to different critical regimes.}
\end{figure}

The next step is to analyse the behaviour of the order parameters.
Continuous phase transitions are described by continuous
dependencies of the order parameters on temperature. Alternatively, if
the function $m_1(T)$ has a gap, this signals that there is a jump
between two different states of the system, which we will associate
with a first order phase transition. As an example, in Fig.
\ref{fig3} we present order parameter dependencies on reduced
temperature $\tau=T/T_c$ for fixed value $\lambda=3.8$ and various
values of $r$. \footnote{Hereafter we are using the value
$\lambda=3.8$ to illustrate typical properties of the system, which
remain qualitatively the same throughout the region $3<\lambda<5$.}

Qualitatively similar behaviour is observed in the whole region $3<\lambda<5$. It is worth noting that in case $r=0$ there is
only one order parameter, as in the genuine Ising model. As one can see from these plots, $m_2$ does not vanish at criticality and
slowly decays as temperature is rising. The same behaviour was observed on the complete graph \cite{Krasnytska16}.

\begin{figure}
    \includegraphics[width=\columnwidth]{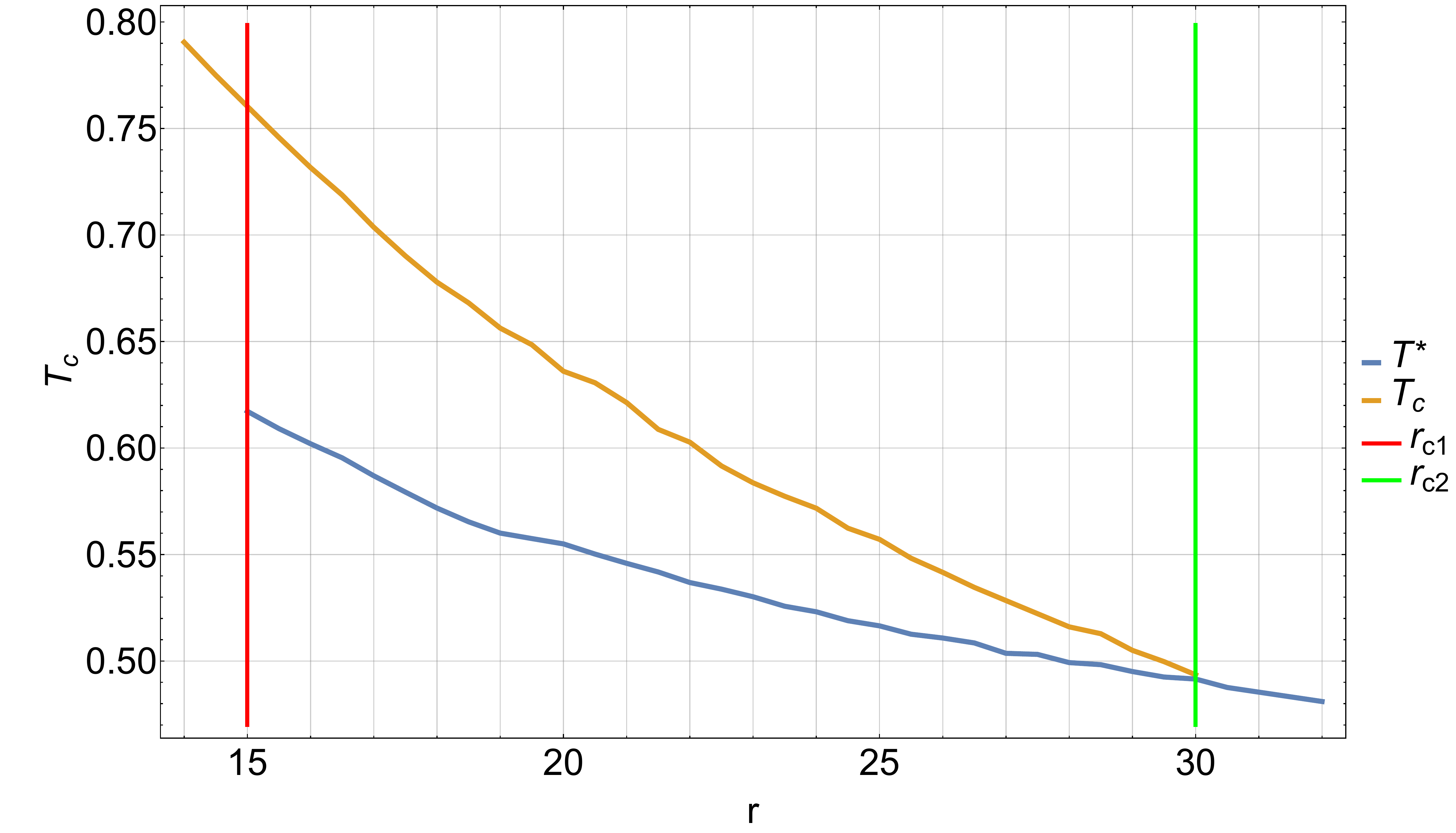}
    \caption{\label{fig3_2} Phase transitions temperatures $T^*$ and $T_c$ as a function $r$ for fixed $\lambda=3.8$. Two solid lines represent first and second order phase transition temperatures; two dashed vertical lines shows marginal values of $r$ and limit the region of the coexistence of two phase transitions.}
\end{figure}

The system undergoes a second order phase transition for small amounts of invisible states. While for the large numbers of invisible states
the phase transition is discontinuous. However, there is a region where two transitions occur at different temperatures:
at lower temperature $T^*$ there is a jump in the order parameter (which we associate with first order phase transitions), and, later,
at higher temperature remaining ordering completely vanishes. Similar behaviour was previously observed in Ref. \cite{Krasnytska16} for the
complete graph, but in the region $1\leq q <2$ only, while limiting case $q=2$ showed sharp distinction between different orders regimes. Topological disorder changes 
critical behaviour. Even in the Ising case, it is characterised
by two marginal values $r_{c1}$  and $r_{c2}$. In Fig. \ref{fig3_2} we show phase diagram in $(T,r)-$plane for the fixed value $\lambda=3.8$. Lower (blue) and upper (yellow) lines represent first and second order phase transition lines respectively.

The second order phase transition line with the first order phase transition line divide $(T,r)-$plane into three regions.
Below the lower (blue) line, the system is in a ordered state, while above the upper (yellow) line - the state is fully disordered.
In the region between the lines, the system is characterised by residual ordering. Therefore, at $r_{c2}$ and $T_c$ these
three phases coincide, making this point tricritical. Two vertical lines mark marginal values $r_{c1}$ and $r_{c1}$, or equivalently the region, where two phase transitions coexist. For each $\lambda$ value there are two marginal values $r_{c1}(\lambda)$ and $r_{c2}(\lambda)$. These two values divide the $(r,\lambda)-$plane into three regions with different critical behaviours.

\begin{figure}
    \includegraphics[width=\columnwidth]{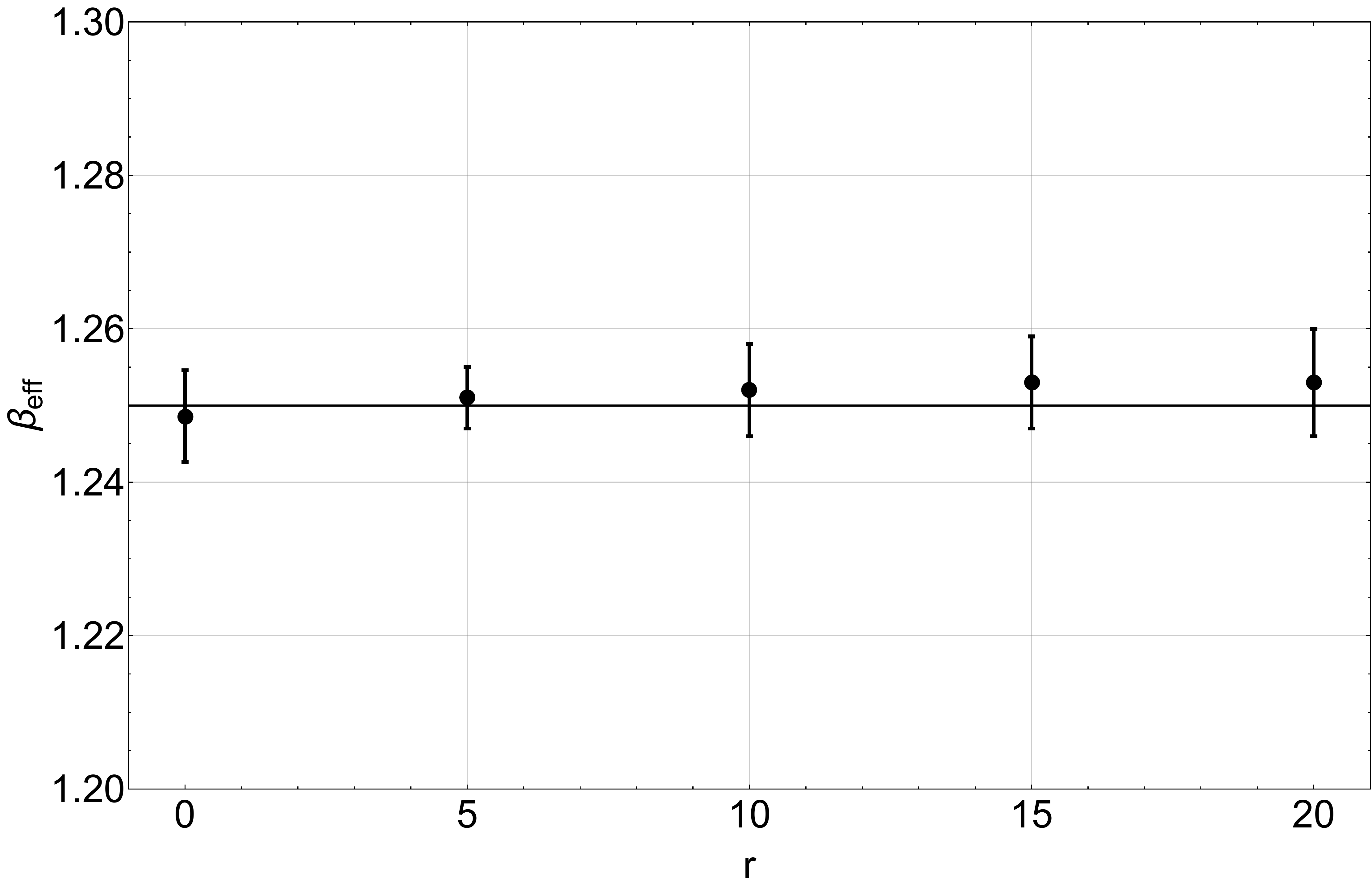}
    \caption{\label{fig5} Critical exponent $\beta$ as a function of number of invisible states $r$ for fixed $\lambda=3.8$. Solid line shows the exact
    result for the genuine Ising model $\beta=\frac1{\lambda-3}|_{\lambda=3.8}=1.25$,  Eq. (\ref{exp}).}
\end{figure}

The next step is to analyse properties of the second order phase transition. With order parameters as function of temperature, it is easy to find critical exponent $\beta$, which is given by:
\begin{equation}
m_1\sim\left(\frac{T_c-T}{T_c}\right)^\beta.
\label{17}
\end{equation}
Since we are minimising free energy numerically, the only way for us
to proceed with the definition (\ref{17}) is to fit obtained values
$m_1(T)$. Because critical exponents are only defined at the
critical temperature, from the fit we only obtain effective
value $\beta_{\rm eff}$. In Fig \ref{fig5} we show critical exponent
$\beta_{\rm eff}$ for different values of $r$. For the genuine Ising
case in the region we are interested in, critical exponents are
$\lambda-$dependent. Analytical results yield
\cite{Leone02,Dorogovtsev02,Palchykov10}:
\begin{equation}\label{exp}
\beta(\lambda)=1/(\lambda-3)\, .
 \end{equation}
In the plot we consider $\lambda=3.8$, thus the theoretical prediction is  $\beta(3.8)=1.25$. This value is shown by a solid horizontal line. We can see that regardless of $r$, a second order phase transition is characterised by the same critical exponent. A slight tendency to increase is due to the fact that the 
effective value of critical exponent is strongly dependent on the region we use for fitting. The smaller the region is, the better fit the first order parameter dependency 
on temperature can be explained by a single power-law \ref{17}. However, with the increase of $r$, the region has to become even smaller, making it much harder to perform numerical calculations very close to the critical temperature.

\begin{figure}
    \includegraphics[width=\columnwidth]{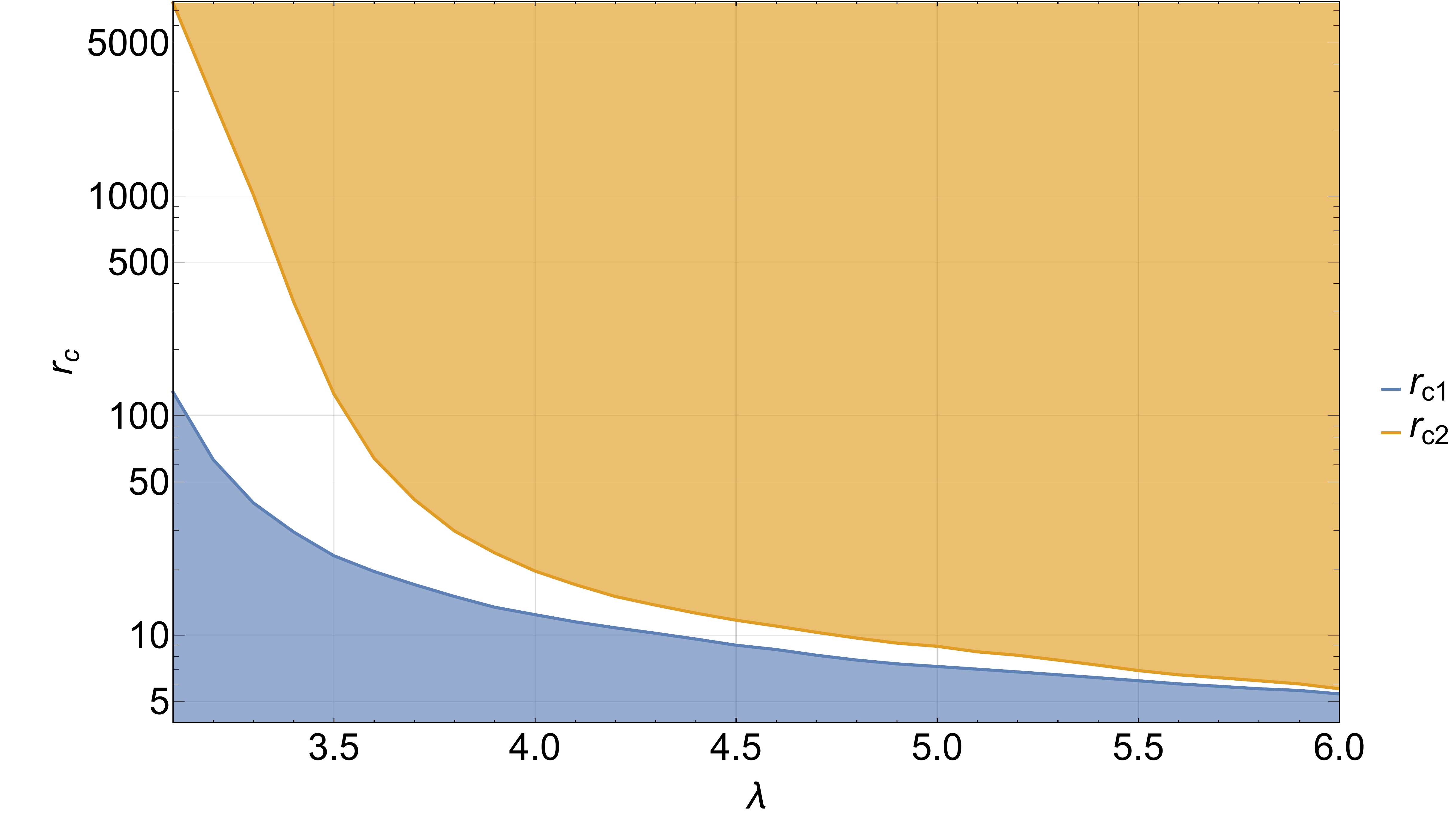}
    \caption{\label{fig4} Phase diagram of the Ising model with invisible states. Three regions, presented here, differ in critical behaviour. In the
    lower (blue) region system possesses only second order phase transition; in the region in-between the lines there are both first and second order phase
    transitions at different temperatures; in upper region (yellow) only the first order phase transition occurs.}
\end{figure}

In Fig. \ref{fig4} the phase diagram in the $(r,\lambda)-$plane is shown. It is characterised by two lines $r_{c1}(\lambda)$ and $r_{c2}(\lambda)$. Below the first one,
with temperature raising, the order parameter continuously changes until it vanishes. Above the $r_{c2}(\lambda)$ line, only the first order phase transition
occurs, meaning that as temperature increases, the order parameter decreases, and at $T_c$ it abruptly drops to zero (see Fig. \ref{fig3} for $r=30$ case).
The system undergoes two phase transitions in the region between the lines. At $T^{*}<T_c$ first order phase transition occur, and there is a jump between two non-zero values of $m_1$. Then, at the second order phase transition temperature $T_c$ first order parameter vanishes.

For fixed value of $\lambda$ and $r$ below the $r_{c1}(\lambda)$ line, $m_1$ depends on temperature continuously. Then the jump in the
order parameter appears. In the region  $r_{c1}(\lambda)<r\leq r_{c2}(\lambda)$ discontinuity grows with $r$ and first order phase transition approaches
second order phase transition $T^{*}\to T_c$. When the $r_{c2}(\lambda)$ line is crossed,
two critical temperatures coincide and only first order phase transition remains, while the residual order parameter value is zero. Note that
in the region $\lambda>4$ the order parameter behaviour close to the second order phase transition temperature is superlinear ($1/\beta=(\lambda-3)$ is
larger than one), which makes distinguishing between first and second order phase transitions even harder. Being evaluated numerically,
$r_{c1}(\lambda)$ and $r_{c2}(\lambda)$ do not cross at $ r\simeq 3.62, \, \lambda=5$, as expected from the analysis of the Ising model with
invisible states on the complete graph \cite{Krasnytska16}.

\section{Conclusions}
\label{IV} Universality is one of the key principles of modern
statistical physics. For lattice systems, critical properties are
defined by space dimensionality, range of interaction and symmetries
of the order parameter. For systems on scale-free networks, degree
distribution exponent plays similar role. In addition, invisible
states have been shown to influence the universality class with only
changing entropic contribution to the free energy
\cite{Sarkanych17,Sarkanych18}. As was shown in this paper, with
these two mechanisms together, the Ising model exhibits non-trivial
properties. The critical temperature is a function of both of these
parameters. The phase diagram, in the region $3\leq\lambda\leq5$, now is
divided into three domains with different critical behaviour: for
$r\leq r_{c1}(\lambda)$ the order parameter depends on the temperature
continuously, meaning that a second order phase transition
occurs; $r_{c1}(\lambda)<r\leq r_{c2}(\lambda)$ at lower temperature
$T^*$ the system undergoes a first order phase transition between two
ordered phases, while at higher temperature $T_c$ second order phase
transitions take place; finally for $r> r_{c2}(\lambda)$ only first
order phase transition remains. This kind of behaviour was earlier
reported for the Potts model with invisible states when $1\leq q<2$.
Here we observe it even in the Ising case $q=2$. We also show, that
adding invisible states does not change the values of the
critical exponents in the region where the second order phase
transition exist.

\section*{Acknowledgement}
We would like to thank Bertrand Berche, Yurij Holovatch and Ralph Kenna for fruitful discussions and useful comments.

\end{document}